\def\Lie{\hbox{\it \char'44}\!}  
\documentstyle[12pt]{article}
\begin{document}

\begin{flushright}
gr-qc/9502037\\
UMDGR-95-100\\
\end{flushright}

\centerline{\Large \bf Spinor one-forms as gravitational potentials}
\vskip 12mm
\centerline{Roh Suan Tung ${}^{\dag}$
  \footnote{Email : m792001@joule.phy.ncu.edu.tw}
       and  Ted Jacobson  ${}^{\ddag}$
  \footnote{Email : jacobson@umdhep.umd.edu} }
\vskip 3mm

\centerline{ ${}^{\dag}$  Department of Physics,
                          National Central University,
                          Chung-Li 32054, Taiwan    }
\centerline{ ${}^{\ddag}$ Department of Physics,
                          University of Maryland,
                          College Park, MD 20742, USA}
\vskip 12mm
\begin{abstract}
General relativity is derived from an action which is
quadratic in the covariant derivative of certain
spinor one-form gravitational potentials.
Either a pair of 2-component spinor one-forms or
a single Dirac spinor one-form can be employed.
The metric is a quadratic function of these spinor one-forms.
In the 2-component spinor formulation the action
differs from the usual chiral action for general relativity
by a total differential.
In the Dirac spinor formulation the action is
the real part of the former one.
The Hamiltonian is related to the ones in positive energy proofs
and spinorial quasilocal mass constructions.
\\ \hbox{} \\ PACS number(s): 04.20
\end{abstract}

\vskip 12mm
We present an action for general relativity, in which the
gravitational variables are
spinor one-form fields and the spin connection.
The action is quadratic in the covariant exterior
derivative of the spinor one-form fields.
If these covariant derivatives are thought of
as the gravitational field strength,
the action is similar to the Yang-Mills action,
although the quadratic function is different and one varies not only
the connection but the spinor one-forms as well.
The Einstein equation is succinctly expressed in terms of
covariant exterior derivatives of the spinor one-form fields.

This formulation is found by a re-examination of
a recent work on a quadratic spinor lagrangian for general
relativity \cite{NT} in a self-dual formulation.
The basic variables of our action are
two SL(2,C) spinor one-forms $\varphi^A, \chi^A$,
which can be realized by treating a tetrad one-form
$\theta^{AA'}$ as two SL(2,C) unprimed spinor one-forms,
$\varphi^A=\theta^{A0'}$ and $\chi^A=\theta^{A1'}$.
A simple {\it quadratic spinor} action for the
spinor one-forms is presented. The formulation is entirely `chiral'
in the sense that it involves only `unprimed' spinor one-forms
$\varphi^A$, $\chi^A$ and the `unprimed' connection one-form
$\omega^{AB}$ as the variables.
The Hamiltonian is
related to the one used in positive energy proofs and quasilocal
mass constructions.
It is shown that alternatively, by combining the two
spinor one-forms into a Dirac spinor one-form
$\Psi=(\varphi_A,\overline\chi^{A'})$,
the same formulation can be derived
from a real valued quadratic spinor action where the variables are
a Dirac spinor one-form and an
SO(3,1) connection one-form.

We begin with the chiral Lagrangian for GR,
which is a 4-form constructed from the tetrad 1-form
$\theta^{AA'}$ and the curvature 2-form
$R_{AB}=d\omega_{AB}+\omega_{A}{}^{C}\wedge\omega_{CB} $
of the spin connection $\omega^{AB}$:
\footnote{The upper case latin letters denote two component spinor
indices, which are raised and lowered with the antisymmetric spinor
$\epsilon_{AB}$ together with its inverse and their conjugates
according to the conventions $\lambda^A:=\epsilon^{AB}\lambda_B$,
$\mu_B:=\mu^A\epsilon_{AB}$, etc. \cite{PR} }
\begin{equation}
{\cal L}_1=
    2 {\rm i}\;\theta^{AA'}\wedge\theta^{B}{}_{A'} \wedge R_{AB}.
\label{L1}
\end{equation}
Let us denote by ${\cal D}$ the covariant exterior differential
acting only on unprimed spinor indices,
${\cal D}\theta^{AA'}=d\theta^{AA'}+\omega^{AB}\wedge\theta_B{}^{A'}$ .
Then $ {\cal D}^2\theta^{AA'}=R^{AB}\wedge\theta_B{}^{A'}$. Therefore,
\begin{equation}
{\cal L}_1={\cal L}_2
  +d(2 {\rm i}\;\theta^{AA'}\wedge {\cal D}\theta_{AA'}),
\label{L1L2}
\end{equation}
where the Lagrangian ${\cal L}_2$ is defined by
\begin{equation}
{\cal L}_2=
    - 2 {\rm i}\; {\cal D}\theta^{AA'}\wedge {\cal D}\theta_{AA'}.
\label{L2}
\end{equation}
The Lagrangian ${\cal L}_2$ can be used to define the action for GR.
Note that while ${\cal L}_1$ is invariant under primed
spin transformations, ${\cal L}_2$ is {\it not} invariant.
However, (\ref{L1L2}) shows that ${\cal L}_2$ changes only
by an exact form under such transformations.

It is natural to eliminate the primed spinor indices
in (\ref{L2}) by introducing a pair of spinor one-forms
$\chi^A, \varphi^A$ defined by
\begin{equation}
 \chi^A  = \theta^{AA'} \iota_{A'} = \theta^{A1'}
\qquad {\rm and}
\end{equation}
\begin{equation}
\varphi^A  = \theta^{AA'}  o_{A'} = \theta^{A0'},
\end{equation}
where $\iota_{A'}, o_{A'}$ is
any pair of spinor fields normalized
such that $\iota^{A'} o_{A'}=1$ (a {\it spin-frame} \cite{PR}).
Conversely, the tetrad can be expressed in terms of the
spinor 1-forms as
\begin{equation}
\theta^{AA'} = \varphi^A \iota^{A'} - \chi^A o^{A'} . \label{eq:c}
 \label{eq:tetrad}
\end{equation}
The metric is then given by
\begin{equation}
g=\theta^{AA'}\otimes\theta_{AA'}
   =\chi^A\otimes\varphi_A + \varphi_A\otimes\chi^A,
\label{g2}
\end{equation}
and $2 \chi^{(A} \wedge \varphi^{B)}:=\Sigma^{AB}$
is the (anti)self-dual 2-form \cite{CDJM}.

In terms of the spinor 1-forms $\chi^A$ and $\varphi^A$
the action $S_2=\int{\cal L}_2$ takes the form
\begin{equation}
S_2[ \chi^A, \varphi^A,  \omega^{AB}]
=4 {\rm i}
\int {\cal D}\varphi^A\wedge {\cal D}\chi_A .
 \label{eq:action}
\end{equation}
The variation of the action (\ref{eq:action})
with $\chi^A$ and $\varphi^A$ held fixed on the boundary
gives the following field equations:
\begin{eqnarray}
 && {\cal D}^2\varphi^A = 0, \qquad
    {\cal D}^2\chi^A = 0,        \label{feqnsa}  \\
 && {\cal D}(\chi_{(A}\wedge\varphi_{B)})=0 .
\label{feqnsb}
\end{eqnarray}
With the tetrad defined in (\ref{eq:tetrad}), the first two field
equations reduce to the complex Einstein equation,
\begin{eqnarray}
 && {\cal D}^2\varphi^A = R^{AB}\wedge\varphi_B
     =-R^{AB}\wedge\theta_{BB'} o^{B'}=0
     \nonumber        \\
 && {\cal D}^2\chi^A=R^{AB}\wedge\chi_B
     =-R^{AB}\wedge\theta_{BB'}\iota^{B'}=0,
\end{eqnarray}
and the third field equation
\begin{equation}
 2 {\cal D}(\chi_{(A}\wedge\varphi_{B)})={\cal D} \Sigma_{AB}
             ={\cal D}(\theta_A{}^{A'}\wedge\theta_{BA'})=0
\label{notorsion}
\end{equation}
states that torsion vanishes and implies that
$\omega_{AB}\epsilon_{A'B'}$ is the
(anti)self-dual part of the connection
determined by the tetrad, which is a function of the spinor 1-forms
$\chi^A, \varphi^A$.
If desired, one can substitute the solution
$\omega_{AB}(\chi, \phi)$ of (\ref{feqnsb}) back into the
action $S_2$ to obtain a {\it second order} action that is
a functional of only the pair of spinor 1-forms.

\setlength{\arraycolsep}{0.5mm}
The Hamiltonian can be constructed \cite{N} from
the action by foliating the spacetime by constant $t$
spacelike hypersurfaces, choosing a
timelike vector field $\xi$ such that $i_\xi dt = 1$,
 and splitting the action:
\begin{eqnarray}
S_2&=&\int {\cal L}_2 = \int dt\wedge i_\xi {\cal L}_2
                      = \int dt \int i_\xi {\cal L}_2 \nonumber\\
&=&4 {\rm i}\int dt\int\Lie_\xi\varphi^A\wedge {\cal D}\chi_A
              +{\cal D}\varphi^A\wedge\Lie_\xi\chi_A  \nonumber \\
 &&           -[ {\cal D} i_\xi\varphi^A\wedge {\cal D}\chi_A
                  + {\cal D}\varphi^A\wedge {\cal D} i_\xi\chi_A
           +i_\xi\omega^{AB}
                 {\cal D}(\varphi_A\wedge\chi_B)]\nonumber\\
&=&\int dt\int 4 {\rm i}\, (\Lie_\xi\varphi^A\wedge{\cal D}\chi_A
           + {\cal D}\varphi^A\wedge\Lie_\xi\chi_A ) -{\cal H} .
\end{eqnarray}
This procedure yields a 4-covariant Hamiltonian 3-form:
\begin{eqnarray}
{\cal H}
  &=& 4 {\rm i}\; [   {\cal D} i_\xi\varphi^A\wedge {\cal D}\chi_A
       + {\cal D}\varphi^A\wedge {\cal D} i_\xi\chi_A
       +i_\xi\omega^{AB}
             {\cal D}(\varphi_A\wedge\chi_B) ] \nonumber\\
  &=& 4 {\rm i}\; [  i_\xi\varphi_A {\cal D}^2\chi^A
       - i_\xi\chi_A  {\cal D}^2\varphi^A +i_\xi\omega^{AB}
        {\cal D}(\varphi_A\wedge\chi_B) + d B ] \nonumber  \\
  &=& 4 {\rm i}\; [ i_\xi\varphi_A R^{AB}\wedge\chi_B
          - i_\xi\chi_A R^{AB}\wedge\varphi_B
          +i_\xi\omega^{AB} {\cal D}(\varphi_A\wedge\chi_B)
          + d B ] \nonumber  \\
  &=& 4 {\rm i}\; [ N^{AA'} R_{AB}\wedge\theta^{BB'} \epsilon_{A'B'}
          +i_\xi\omega^{AB} {\cal D}(\varphi_A\wedge\chi_B)
          + d B ] ,
\end{eqnarray}
where  $B= i_\xi\varphi^A {\cal D}\chi_A
          +i_\xi\chi_A {\cal D}\varphi^A$
and
$N^{AA'}=i_\xi \varphi^A \iota^{A'}-i_\xi \chi^A o^{A'}
=i_\xi\theta^{AA'}$ is the evolution vector.
The Hamiltonian 3-form differs from the usual (linear in complex
Einstein tensor) Hamiltonian for self-dual gravity \cite{A}  by a
total differential.
It has asymptotically flat fall off of $O(1/r^4)$, and
the variation of the Hamiltonian will have an $O(1/r^3)$ boundary term
which vanishes asymptotically.
Therefore the Hamiltonian is well-defined \cite{RT} .

With the field equations satisfied,
the Hamiltonian 3-form reduces to an exact differential.
Integration yields an integral over a 2-surface
\begin{equation}
 E=\int {\cal H} \mid_{\rm S}
    =4 {\rm i} \oint i_\xi\varphi^A {\cal D}\chi_A
          + i_\xi\chi_A {\cal D}\varphi^A
   = 4 {\rm i} \oint N^{AA'} {\cal D} \theta_{AA'} ,
 \label{eq:energy}
\end{equation}
which determines the energy within a 2-surface.
This is comparable with the usual spinorial quasilocal energy
constructions \cite{QLE}
\begin{equation}
 E^\prime =-4 {\rm i} \oint
\overline{\lambda}_{A'} {\cal D}\lambda_A\wedge \theta^{AA'}.
\end{equation}
In (16) the spinors $\lambda^{A}$, $\overline{\lambda}{}^{A'}$
define the null evolution vector,
whereas the spinor 1-forms
in the energy expression (\ref{eq:energy})
are field variables in the Lagrangian, and the evolution vector
$N^{AA'}$ need not be null.
When the evolution vector is null and thus expressed as
$N^{AA'}=\lambda^{A} \overline{\lambda}{}^{A'}$,
the difference of the two energy expressions $E$ and $E'$
is given by
\begin{equation}
E-E'= -4 {\rm i} \oint
    \lambda^A d\overline{\lambda}{}^{A'}
          \wedge\theta_{AA'}.
\end{equation}
For asymptotically flat spacetime, this difference
vanishes at infinity, since the integrand becomes a total
derivative $d\Bigl(\overline{\lambda}{}^{A'}
(\lambda^A \theta_{AA'})_\infty \Bigr)$.
Therefore the two energy expressions
yield the same value at infinity.

Now we show that,
by combining the two spinor 1-form fields into a Dirac spinor
$\Psi=(\varphi_A,\overline{\chi}^{A'})$ in the Weyl representation,
the action (\ref{eq:action})
can be alternatively written as \cite{NT}
\begin{equation}
 S_{\Psi}[\overline\Psi, \Psi, \omega^{ab} ]=
 \int{\cal L}_\Psi = 2
 \int  \overline{D\Psi} \wedge \gamma_5 D\Psi.
 \label{eq:iaction}
\end{equation}
In contrast to the point of view in \cite{NT},
$\Psi$ here is a fundamental field variable to be freely varied,
whereas in \cite{NT} it was constructed from the tetrad and
a fixed, normalized Dirac spinor field. (This is a distinction
without a difference however, since the tetrad variations
of \cite{NT} result in free variations of $\Psi$,
and no impression of the fixed normalized
Dirac spinor survives.)
The {\it real} SO(3,1) connection 1-form
$\omega^{ab}=\omega^{AB}\epsilon^{A'B'}+
\overline{\omega}^{A'B'}\epsilon^{AB}$
consists of the unprimed self-dual connection and its conjugate.

In the Weyl representation,
the covariant derivative of the Dirac spinor 1-forms is given by
\begin{equation}
\overline{D\Psi}=
({\cal D}\chi^A\quad\overline{{\cal D}}\overline{\varphi}_{A'}),
\qquad
D\Psi={{\cal D}\varphi_A
\choose \overline{{\cal D}}\overline{\chi}^{A'}},
\qquad
\end{equation}
and $\gamma_5:=\gamma_0\gamma_1\gamma_2\gamma_3=
{\rm i} \; {\rm diag}(-1,-1,1,1)$.
The Lagrangian in (\ref{eq:iaction}) can thus
be split into unprimed and primed parts:
\begin{equation}
 {\cal L}_\Psi
  = 2 {\rm i}\;{\cal D}\varphi^A\wedge {\cal D}\chi_A
   -2 {\rm i}\;\overline{{\cal D}}\overline{\varphi}^{A'}\wedge
     \overline{{\cal D}}\overline{\chi}_{A'} ,
\label{LPsi}
\end{equation}
where the second term is just the conjugate of the first term.
The action $S_{\Psi}$ in (\ref{eq:iaction})
is therefore just the real part of the
action $S_2$ (\ref{eq:action}).

Now a key observation is that $S_2[z]$ is a holomorphic
function of $z$, where
$z:=(\chi^{A},\varphi^{A},\omega^{AB})$.
Therefore, just as for analytic functions
of a finite number of variables, the derivative
 $\delta S[z]/\delta z$ vanishes if and only if
the derivative of the real (or imaginary) part of
$S[z]$ vanishes. Thus, as far as the equations of motion
are concerned, the action $S_{\Psi}$ (\ref{eq:iaction})
is {\it equivalent} to the original
chiral action $S_2$ (\ref{eq:action}).

The metric (\ref{g2})
is given in terms of the Dirac spinor 1-form by:
\begin{equation}
g=\overline\Psi\otimes_{\!S} (1+{\rm i} \gamma_5)\Psi,
\end{equation}
where $\otimes_{\!S}$ denotes the symmetrized tensor product
($X\otimes_{\!S} Y\equiv (X\otimes Y + Y\otimes X)/2$.)
Note that $\overline\Psi\otimes_{\!S} \Psi$ and
$\overline\Psi\otimes_{\!S} \gamma_5\Psi$
are the real and imaginary parts of the complex metric respectively.
When the metric is real it is therefore given by
\begin{equation}
g=\overline\Psi\otimes_{\!S} \Psi,
\end{equation}
and the reality condition for the metric is
$\overline\Psi\otimes_{\!S} \gamma_5\Psi=0$.

When the metric is real, the Lagrangian (\ref{LPsi})
is the usual, real, Hilbert-Palatini Lagrangian plus a
total differential. This can be seen as follows.
In terms of an arbitrary spin frame and the corresponding
tetrad (\ref{eq:tetrad})
the Lagrangian (\ref{LPsi}) is written in terms of the
Lagrangian (\ref{L1}) as
\begin{eqnarray}
  {\cal L}_\Psi
 &=&\textstyle\frac{1}{2}({\cal L}_1 + \overline{\cal L}_1)+d(...)
  \nonumber\\
 &=& -{\rm i}\;\theta^{AA'}\wedge\theta^{BB'}
    \wedge R_{AB} \epsilon_{A'B'}
 +{\rm i}\;\overline{\theta^{AA'}}\wedge\overline{\theta^{BB'}}\wedge
    \overline{R}_{A'B'} \epsilon_{AB} +d(...)
\end{eqnarray}
If the metric is real one can always find a spin frame
such that $\theta^{AA'}$ is real, in which case
$\overline{\theta^{AA'}}=\theta^{AA'}$,
and the Lagrangian becomes
\begin{eqnarray}
 {\cal L}_\Psi
 &=& \theta^{AA'}\wedge\theta^{BB'}\wedge (-{\rm i}\; R_{AB}
     \epsilon_{A'B'}+
 {\rm i}\; \overline{R}_{A'B'} \epsilon_{AB})  +d(...) \nonumber\\
 &=& \theta^{a}\wedge\theta^{b}\wedge\ast R_{ab}+d(...),
\end{eqnarray}
which is the usual real Hilbert-Palatini Lagrangian
plus a total differential.

The Dirac spinor formulation is thus
``more real" than that of the
usual chiral action, or the action (\ref{eq:action}), because
($i$) the action (\ref{eq:iaction})
is always real and ($ii$) it is the real
Hilbert-Palatini action when the metric is real, without
requiring that the connection satisfy its equation of motion.

\vskip 1cm
We would like to thank Prof. J. M. Nester for helpful discussions.
RST was supported in part by NSC 84-2112-M-008-004,
and TJ by NSF grant PHY94-13253.
RST would like to thank Department of Physics, University of Maryland
at College Park for hospitality while some of this work was carried out.

\vskip 1cm
\eject

\end{document}